\documentclass[prb,twocolumn,showpacs,preprintnumbers,amsmath,amssymb]{revtex4}

\usepackage{graphicx}
\usepackage{dcolumn}
\usepackage{bm}
\usepackage{mathrsfs}
\usepackage{float}
\usepackage{wrapfig}
\usepackage{color}

\begin{document}


\title{Transport criticality at Mott transition in a triangular lattice Hubbard model}

\author{Toshihiro Sato}
\author{Kazumasa Hattori }
\author{Hirokazu Tsunetsugu}

\affiliation{
Institute for Solid State Physics, University of Tokyo, 5-1-5 Kashiwanoha, Kashiwa, Chiba 277-8581, Japan}

\date{\today}

\begin{abstract}

We study electric transport near the Mott metal-insulator transition in a triangular-lattice Hubbard model at half filling.
We calculate optical conductivity $\sigma(\omega)$ based on a cellular dynamical mean field theory including vertex corrections
inside the cluster.
Near the Mott critical end point, a Drude analysis in the metallic region suggests that the change in the Drude weight
is important rather than that in the transport scattering rate for the Mott transition.
In the insulating region, there emerges an ``ingap" peak in $\sigma(\omega)$ at low $\omega$ near the Mott transition,
and this smoothly connects to the Drude peak in the metallic region with decreasing Coulomb repulsion.
We find that the weight of these peaks exhibits a power-law behavior upon controlling Coulomb repulsion
at the critical temperature.
The obtained critical exponent suggests that conductivity does not correspond to magnetization or energy density of the
Ising universality class in contrast to several previous works.

\end{abstract}

\maketitle

\section{\label{sec:Int}Introduction}

The Mott metal-insulator transition is one of fascinating phenomena in strongly correlated electronic systems.
It is realized in two ways; bandwidth-control at fixed electron density and electron-density-control at fixed
Coulomb repulsion and bare bandwidth.
In this paper, we focus on the former type.
The Mott transition occurs when the Coulomb repulsion predominates over the electronic kinetic energy
and it is not accompanied by magnetic transition in the original sense.
This takes place inside the paramagnetic phase at finite temperature and magnetic frustration plays an important role for its realization.
Its phase diagram in the parameter space of temperature and (chemical) pressure has insulating and
metallic phases which are separated by a line of first-order Mott transition at low temperature.
This phase boundary terminates at a
critical end point, namely, the Mott critical end point.
This behavior has been observed in several materials such as a classic three-dimensional Cr-doped V$_2$O$_3$ \cite{MT-exp-1}
and quasi-two-dimensional $\kappa$-type organic compounds.\cite{MI-tri-1}
This is similar to the liquid-gas transition in classical liquids~\cite{LC-tri-1} and critical behaviors
are expected in various properties around the Mott critical end point.

In the Mott transition, double occupancy (doublon density) $d=\langle n_{i\uparrow}n_{i\downarrow}\rangle$
plays the role of order parameter,\cite{docc-op} where $n_{i\sigma}$ denotes the electron density operator
at site $i$ with spin $\sigma$.
Recently, the existence of critical behaviors of double occupancy near the Mott critical end point has been reported
in some theoretical works.
Its criticality, ``Mott criticality," is analyzed with respect to two control parameters.
One is simply temperature, and the other is the conjugate field of the order parameter.
In our case, the order parameter is double occupancy, and this appears in the Hubbard Hamiltonian 
with multiplied by Coulomb repulsion $U$.
Therefore, $U$ plays the role of the conjugate field in the theory of Mott transition.
In experiments, however, it is not easy to control Coulomb interaction directly.
It is standard to control pressure, instead, since this changes the bandwidth $W$ of the material and therefore
control effectively the ratio $U/W$.
One should note that applying pressure corresponds to decreasing effective $U$.
A Ginzburg-Landau analysis \cite{MI-crit-1} or an scaling analysis based on the dynamical mean field calculations \cite{MTC-dcc-DMFT-1,docc-scaling}
have suggested
that double occupancy shows the same scaling behavior as the Ising order parameter and Coulomb repulsion corresponds
to magnetic field in the Ising model.
This is equivalent to the liquid-gas transition in classical liquids.
However, the Mott criticality in other properties is not well understood and transport properties are particularly important
among them.
This is the main issue of this paper and we will report our numerical study on the Mott criticality in electronic transport
of a triangular-lattice Hubbard model.

To study electric transport in strong correlated electronic systems, not only dc-electric conductivity but ac response,
i.e., optical conductivity, provide useful information on the charge dynamics, in particular, effective mass,
transport scattering process and electric structure.
Optical conductivity experiments for several $\kappa$-type organic compounds have examined so far.
Insulating regular triangular lattice compound $\kappa$-(ET)$_{2}$Cu$_{2}$(CN)$_{3}$ shows spin liquid behaviors
with no magnetic long-range order.\cite{MI-tri-1,SL-1,SL-2,SL-3}
Its optical conductivity does not show a clear gap but decays smoothly toward zero frequency.\cite{OC-1}
In contrast, $\kappa$-(ET)$_{2}$Cu[N(CN)$_{2}$]Cl is less frustrated due to a distortion in the triangular structure
and exhibits an antiferromagnetic order.
Its optical conductivity has a clear gap instead.\cite{OC-2,OC-3}

Recently, two experimental studies have reported critical behaviors in dc-electric conductivity $\sigma_{0}$
near the Mott critical end point, but the two conclusions are not consistent to each other. 
Limelette~{\it et al.}~have assumed that $\sigma_{0}$ has the same singularity as order parameter in the Mott transition
and analyzed its singularity in (V$_{\rm 1-x}$Cr$_{\rm x}$)$_2$O$_3$.
Translating the criticality of magnetization with respect to temperature and magnetic field in the Ising universality class,
they expected
\begin{eqnarray}
\sigma_{0}(t=0,p>0)-\sigma_{0}^* &\sim& |p|^{1/\delta},\\
\sigma_{0}(t,p=0)-\sigma_{0}^* &\sim& |t|^{\beta},\\
\frac{\partial \sigma_{0}(t,p)}{\partial p} \Bigr|_{p=0} &\sim&|t|^{-\gamma},
\label{eq:exp-scal}
\end{eqnarray}
where $t=(T-T^*)/T^*$ and $p=(P-P^*)/P^*$.
$(\sigma_{0}^*, T^*, P^*)$ denote the values at the Mott critical end point.
Recall that pressure plays the role of a conjugate field to $\sigma_{0}$ in their experiment.
The estimated values of the critical exponents are $(\delta, \beta, \gamma)\sim(4.8, 0.33, 1.2)$\cite{MTC-exp-V}
and they agree with the values for the three-dimensional Ising universality class.
A similar analysis was carried out by Kagawa~{\it et al.} for the quasi-two-dimensional anisotropic triangular lattice compound $\kappa$-(ET)$_{2}$Cu[N(CN)$_{2}$]Cl.
However, the estimated critical exponents are now $(\delta, \beta, \gamma)\sim(2, 1, 1),$\cite{MTC-exp-kapp}
different from the values obtained by Limelette~{\it et al.}

Theoretical understanding of these inconsistent results is still controversial.
Imada~{\it et al.} argued that critical exponents by Kagawa~{\it et al.} are due to a marginal quantum critical point in the two-dimensional system.
\cite{MI-crit-1,MI-crit-2,MI-crit-3}
Papanikolaou~{\it et al.} proposed another theory that the critical exponents deviate due to large
subleading corrections related to energy density.\cite{MTC-kapptheo-2}
The first numerical study was performed by Sentef~{\it et al.} based on the cellular dynamical mean field theory
and it was claimed that the calculated imaginary-time Green's function shows the same criticality as the experimental
data of $\sigma_{0}$.\cite{MTC-kapptheo-3}

Among these theories, the first two are semi-phenomenological approaches,
and in the last one the comparison was not made for the identical quantity.
Therefore, it is important to study the critical exponent of conductivity directly by numerical calculations.
However, it is very difficult to obtain the precise exponent for conductivity,
since we need to employ some approximate scheme like cellular dynamical mean field theory with a small cluster size.
One important issue is whether the transport critical exponent agrees with the one of the order parameter,
and this is the problem that we study in this paper.
To this end, we calculate these critical exponents within the same approximation.
Of course, the obtained values are not exact critical exponents, due to approximation,
but if they differ to each other, this implies that their exact exponents are also distinct.
We will actually demonstrate that this is the case by using the the cellular dynamical mean field approach in this paper.

To take into account correlation effects and frustrated lattice geometry, we calculate optical conductivity based on the Kubo formula\cite{Kubo}
and the cellular dynamical mean field theory including vertex corrections.
This is one of the first achievements of numerical approach for conductivity, which is a big challenge in numerical computations.

Our paper is organized as follows.
In Sec.~\ref{sec:MM}, we first review our numerical method, the cellular dynamic mean field theory.
Then, we explain how to calculate optical conductivity in the cellular dynamical mean filed theory
including vertex corrections.
In Sec.~\ref{sec:ThC}, we discuss thermodynamic criticality near the Mott critical end point.
In Sec.~\ref{sec:EP}, we examine electronic properties near the Mott critical end point.
In Sec.~\ref{sec:TP}, we present our results of optical conductivity with controlling Coulomb repulsion.
Sec.~\ref{sec:IG} is devoted to the examination of a low-energy small peak in optical conductivity on the insulating side.
In Sec.~\ref{sec:TrC}, we present our result of scaling analysis of optical conductivity.
Sec.~\ref{sec:SD} is summary and discussion.

\section{\label{sec:MM}Model and Method}

\subsection{\label{sec:level1}Cellular dynamical mean field theory}

\begin{figure}
\centering
\includegraphics[width=8cm,bb=0 0 575 525]{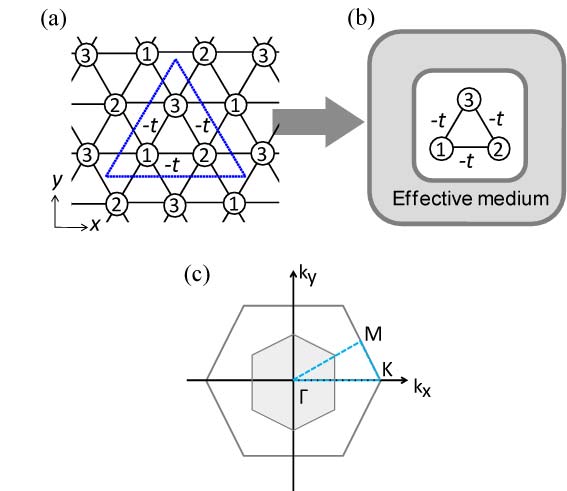}
\caption{(Color online) (a) Isotropic triangular lattice and (b) three-site triangular cluster model employing CDMFT.
(c) First Brillouin zone of an isotropic triangular lattice. Gray zone draws the reduced Brillouin zone for the sublattice
using the three-site triangular cluster.  
}
\label{fig:CDMFT}
\end{figure}

The model we consider in this paper is a single-band Hubbard Hamiltonian on an isotropic triangular lattice at half filling,
\begin{eqnarray}
H=-t\sum_{\langle i,j \rangle,\sigma}c_{i\sigma}^\dagger c_{j\sigma}+U\sum_{i}n_{i\uparrow}n_{i\downarrow}-\mu\sum_{i,\sigma}
c_{i\sigma}^\dagger c_{i\sigma}.
\label{eq:H-1}
\end{eqnarray}
Here, $t$ is the nearest-neighbor hopping amplitude and $U$ is the on-site Coulomb repulsion.
$\mu$ is the chemical potential and is set by tuning electron density to half filling in the case of non bipartite system.
$c_{i\sigma}$ is the electron annihilation operator at site $i$ with spin $\sigma$,
and $n_{i\sigma}=c_{i\sigma}^\dagger c_{i\sigma}$.
In what follows, $U$ and $T$ are in units of $t$.
The energy dispersion of the kinetic term is given by 
\begin{eqnarray}
\epsilon(\mathbf k)=-2t\Bigl(\rm cos \it k_x+\rm 2cos \frac{\it k_x}{2}cos \frac{\sqrt{3} \it k_y}{\rm 2}\Bigr),
\label{eq:ED-1}
\end{eqnarray}
where $\mathbf k$ is the wave vector.

Our primary concern is about Mott transition, and it occurs in situations where $U$ is dominant over kinetic energy,
which is proportional to $t$.
To take into account both strong electronic correlations and geometrical frustration, we use the cellular dynamical mean field theory (CDMFT).\cite{CDMFT}
Considering the lattice structure, we map the Hamiltonian (\ref{eq:H-1}) onto a three-site triangular cluster model coupled to an effective medium $\hat{\mathcal{G}}_{\sigma}(i\omega_{n})$ with Matsubara frequency $i\omega_{n}$
determined self-consistently as shown in Figs.~\ref{fig:CDMFT} (a) and (b).
This self-consistency equation is given by
\begin{eqnarray}
\hat{\mathcal{G}}_{\sigma}(i\omega_{n})&=&\Biggl[\sum_{\mathbf K}\frac{1}{(i\omega_{n}+\mu)\hat{\mathbf 1}-\hat{t}(\mathbf K)-\hat{\Sigma}_{\sigma}(i\omega_{n})}\Biggr]^{-1}\nonumber \\
&+&\hat{\Sigma}_{\sigma}(i\omega_{n}),
\label{eq:H-2}
\end{eqnarray}
where $\mathbf K$ is the wave vector in the reduced Brillouin zone for the sublattice as shown in Fig.~\ref{fig:CDMFT} (c).
$\hat{t}(\mathbf K)$ is the Fourier transformed hopping matrix for the sublattice
and $\hat{\Sigma}_{\sigma}(i\omega_{n})$ is the cluster self-energy.
We compute the cluster Green's function 
$ \bigl[\hat{G}_{\sigma}(\tau)\bigr]_{\alpha \beta} \equiv -\langle T_{\tau}c_{\alpha \sigma}(\tau)c_{\beta \sigma}^\dagger(0)\rangle $
by using the continuous-time quantum Monte Carlo (CTQMC) method based on the strong coupling expansion.~\cite{CTQMC}
Here, $\alpha$ and $\beta$ denote sites in the cluster and $\tau$ is imaginary time.
The cluster self-energy is calculated via the Dyson equation,
\begin{eqnarray}
\hat{\Sigma}_{\sigma}(i\omega_{n})=\hat{\mathcal{G}}_{\sigma}(i\omega_{n})-\hat{G}_{\sigma}^{-1}(i\omega_{n}).
\label{eq:H-3}
\end{eqnarray}
Here, $\hat{\mathcal{G}}_{\sigma}(i\omega_{n})$, $\hat{t}(\mathbf K)$, $ \hat{G}_{\sigma}(i\omega_{n})$
and $\hat{\Sigma}_{\sigma}(i\omega_{n})$ are $3 \times 3$ matrices.
In our CDMFT calculations, we consider only paramagnetic solution.

\begin{figure}
\includegraphics[width=7.5cm,bb=0 0 500 950]{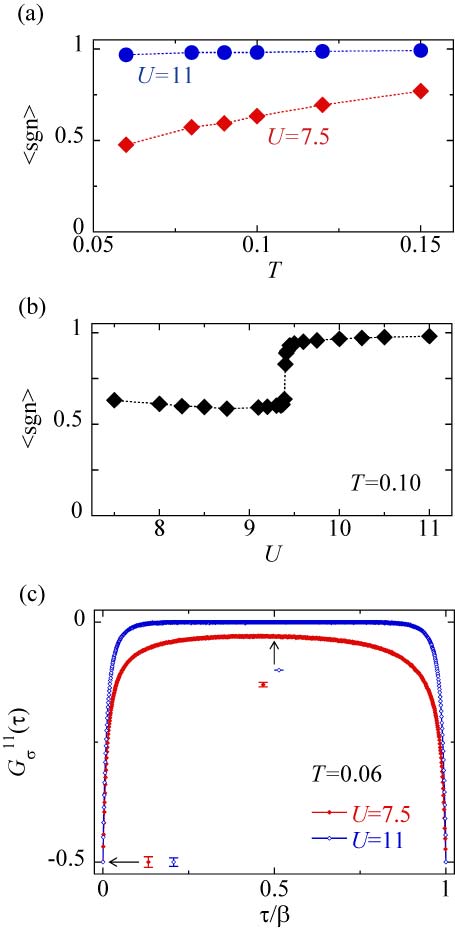}
\caption{(Color online) (a) $T$-dependence of the average sign $\langle sgn \rangle$ for two $U$'s and
(b) $U$-dependence of $\langle sgn \rangle$ at $T=0.10$.
(c) Local Green's function $G_{\sigma}^{11}(\tau)$ at $T=0.06$ for $U=7.5$ and 11.
Error bars are shown at two typical values of $\tau$: $\tau=0$ and $\beta/2$.
}
\label{fig:sgn}
\end{figure}

The negative sign problem in the MC calculations is one of the important issues, particularly on the frustrated lattice.
We have succeeded in reducing it by using molecular orbital bases.
Our cluster has $C_{3v}$ point group symmetry, and electron operators in the molecular orbital basis are
\begin{eqnarray}
a_{1\sigma}&=&\frac{1}{\sqrt{3}}(c_{1\sigma}+c_{2\sigma}+c_{3\sigma}),\nonumber \\
a_{2\sigma}&=&\frac{1}{\sqrt{2}}(c_{1\sigma}-c_{3\sigma}),\\
a_{3\sigma}&=&\frac{1}{\sqrt{6}}(c_{1\sigma}-2c_{2\sigma}+c_{3\sigma}).\nonumber
\label{eq:MB-1}
\end{eqnarray}
In these molecular orbital basis, 
the effective medium and the cluster Green's function can be represented by diagonal $3 \times 3$ matrices.
Negative sign problem is not serious when calculations are performed with these basis functions.
Figures~\ref{fig:sgn} (a) and (b) show the average sign,
\begin{eqnarray}
\langle sgn \rangle=\frac{\sum_{i}p_i}{\sum_{i}|p_i|},
\label{eq:sng-1}
\end{eqnarray}
in the MC calculations on $U$-$T$ parameter space, where $p_i$ is the sign of the weight for MC configurations. 
One can see that $\langle sgn \rangle$ is large even for smaller $U$ and lower $T$, and this guarantees higher-accuracy calculations for various quantities.
For example, we have calculated electron density $n$ by performing $2 \times 10^{7}$ MC sweeps and averaging over 64 samples.
The error $\Delta n$ is small and the typical value is $\Delta n/n \sim 0.003$ for $U=7.5$ and $11$ at $T=0.06$.
In Fig.~\ref{fig:sgn}~(c), we show the diagonal part of the cluster Green's function $G_{\sigma}^{11}(\tau)$ as a function of imaginary time $\tau$,
for two values of $U$.
Its error at $\tau=0$ is $(\Delta G_{\sigma}^{11}/G_{\sigma}^{11})(\tau=0) \sim 0.02$,
and this is larger than the one in the direct measurement $\Delta n/n \sim 0.003$ but is still quite small.
To determine the chemical potential $\mu$, we have used the directly calculated value of $n$ rather than calculating from $G_{\sigma}^{11}$.

After the self-consistency loop of CDMFT converges, we calculate the lattice Green's functions $G_{\sigma}(\mathbf k,i\omega_{n})$
of the triangular lattice from the cluster quantities.
There are several methods for this interpolation, and we use the cumulant method in this study.
It has been shown that the cumulant interpolation works well in a wide range of $U$
from metallic to insulating state.\cite{GF-cal-1,GF-cal,GF-cal-2}
We first introduce and calculate the cluster cumulant
\begin{eqnarray}
\hat{M}_{\sigma}(i\omega_{n})=\frac{1}{(i\omega_{n}+\mu)\hat{\mathbf 1}-\hat{\Sigma}_{\sigma}(i\omega_{n})},
\label{eq:lattice-1}
\end{eqnarray}
and proceed to calculate the cumulant on the original lattice system
\begin{eqnarray}
M_{\sigma}(\mathbf k,i\omega_{n})=\frac{1}{N_{\rm C}}\sum_{\alpha \beta}M^{\alpha \beta}_{\sigma}(i\omega_{n})e^{-i\mathbf k \cdot(\mathbf r_\alpha-\mathbf r_\beta)},
\label{eq:lattice-2}
\end{eqnarray}
where $N_{\rm C}(=3)$ is the cluster size and $\mathbf r_\alpha$ is the real-space position of site $\alpha$.
Then, the lattice Green's function is given as
\begin{eqnarray}
G_{\sigma}(\mathbf k,i\omega_{n})=\frac{1}{M_{\sigma}(\mathbf k,i\omega_{n})-\epsilon(\mathbf k)}.
\label{eq:lattice-3}
\end{eqnarray}

\begin{figure}
\centering
\centerline{\includegraphics[width=5cm,bb=0 0 150 130]{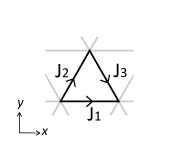}}
\caption{(Color online) Particle current in an isotropic triangular lattice. 
}
\label{fig:current}
\end{figure}

\subsection{\label{sec:level1}Optical conductivity including vertex corrections}

Let us discuss vertex corrections in details.
In the single-site DMFT, the current-current correlation function has no vertex correction 
and conductivity is calculated by simple convolution of the lattice Green's function $G_{\sigma}(\mathbf k,i\omega_{n})$.\cite{OC-DMFT,OC-DMFT-2}
This is because current is a site off-diagonal operator, and this point is different from the case of, for example, spin susceptibility,
where spin density is site diagonal.
Effects of vertex corrections on spin susceptibility have been studied already in the single-site DMFT approach.\cite{SC-DMFT,SC-DMFT-1,SC-DMFT-2,SC-DMFT-3,SC-DMFT-4}
Vertex corrections have finite effects on conductivity in the CDMFT except in some special cases,\cite{OC-CDMFT-TJ-nov-1,OC-CDMFT-TJ-nov-2}
but in some preceding works of the CDMFT, conductivity has been calculated without vertex corrections.
\cite{OC-CDMFT-nov1,OC-CDMFT-nov2}
Their effects were studied particularly for the square lattice cases using dynamical cluster approximation (DCA),
which is a cluster extension of DMFT in the $\mathbf k$-space.\cite{OC-DCA,OC-DCA-2}
To take into account the correlation effects in current response further, we use CDMFT and calculate optical conductivity
including vertex corrections.
This is a big challenge in numerical computations.

We start with explaining our algorithm including the vertex corrections in CDMFT.
Optical conductivity tensor $\sigma_{\nu \nu'}(\omega)$ is generally defined by the Kubo formula~\cite{Kubo} as
\begin{eqnarray}
\sigma_{\nu \nu'}(\omega)=-\frac{ie^2}{\hbar}{\rm \lim_{\mathbf q \rm \rightarrow 0}}\frac{\chi_{\nu \nu'}(\mathbf q,\omega)
-\chi_{\nu \nu'}(\mathbf q,0)}{\omega},
\label{eq:OC-0}
\end{eqnarray}
where $e$ is the elementary charge, $\omega$ is the frequency, and $\mathbf q$ is wave vector.
Here, $\chi_{\nu \nu'}(\mathbf q, \omega)$ is the current-current correlation function,
and is calculated from Fourier transform of the correlation function in real time $t$
\begin{eqnarray}
\chi_{\nu \nu'}(\mathbf q,t)=\langle J_{\nu}(\mathbf q,t)J_{\nu'}(\mathbf q,0)\rangle,
\label{eq:OC-4}
\end{eqnarray}
where $J_{\nu}(\mathbf q,t)$ is the particle current operator in the $\nu(=x,y)$ direction
and is defined as follows,
\begin{eqnarray}
J_{\nu}(\mathbf q, t)&=&-\frac{2t}{N}\sum_{\mathbf r} (\mathbf r)_{\nu} \sum_{\mathbf k, \sigma}
\rm{sin} \bigl[\bigl(\mathbf{k}-\frac{\mathbf q}{2}\bigr)\cdot \mathbf{r}\bigr]\nonumber \\
&\times&\it c_{\mathbf{k}-\frac{\mathbf q}{\rm 2}\it \sigma}^\dagger(\it t)\it c_{\mathbf{k}+\frac{\mathbf q}{\rm 2}\it \sigma}(\it t),
\label{eq:OC-5}
\end{eqnarray}
where $a$ is the lattice unit and $N$ is the total number of sites. 
$\mathbf r$$=$$(a,0)$, $(a/2,\sqrt{3}a/2)$, or $(a/2, -\sqrt{3}a/2)$.
In a triangular lattice system, the particle current is defined on each of the three bonds as shown in Fig.~\ref{fig:current}
and the $x$-component of the net current is given as $J_1+(J_2+J_3)/2$.
In our calculations, we only focus on $\sigma_{xx}(\omega)$ because of $\sigma_{\nu \nu'}(\omega)=\sigma(\omega)\delta_{\nu \nu'}$
in an isotropic triangular lattice.
We denote $\sigma(\omega) \equiv \sigma_{xx}(\omega)$ and $\chi(i\nu_n) \equiv \chi_{xx}(i\nu_n)$ later.
From Eqs.~(\ref{eq:OC-4}) and (\ref{eq:OC-5}), $\chi(t) \equiv \chi(\mathbf q\rightarrow0,t)$
is given by
\begin{eqnarray}
\chi(t)
=\frac{1}{N^2}\sum_{\mathbf k,\mathbf k',\sigma,\sigma'}v_{\mathbf k}^{x}v_{\mathbf k'}^{x'}
\langle c_{\mathbf k \sigma}^\dagger(t)c_{\mathbf k \sigma}(t)c_{\mathbf k' \sigma'}^\dagger(0) c_{\mathbf k' \sigma'}(0)\rangle,\nonumber \\
\label{eq:OC-6}
\end{eqnarray}
where $v_{\mathbf k}^{x}$ is the $x$-component of velocity described as
\begin{eqnarray}
v_{\mathbf k}^x=2ta\Bigl(\rm sin \it{k_x a}+\rm sin \frac{\it{k_x a}}{\rm2}
\rm cos \frac{\sqrt{\rm3}\it{k_y a}}{\rm 2} \Bigr).
\label{eq:OC-7}
\end{eqnarray}

\begin{figure}
\centering
\centerline{\includegraphics[width=8cm,bb=0 0 500 650]{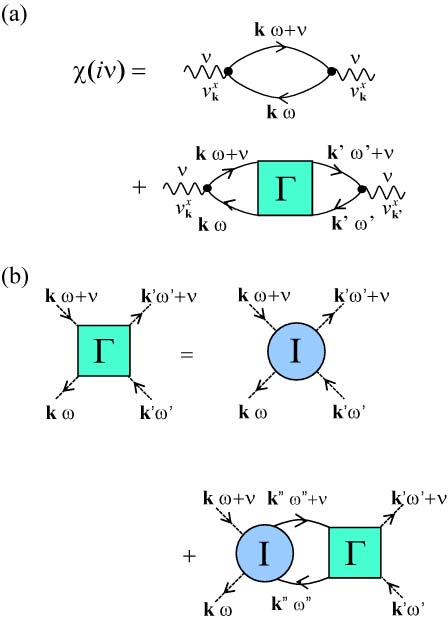}}
\caption{(Color online) Diagrams for (a) the current-current correlation function $\chi(i\nu_n)$ in Eq.~(\ref{eq:OC-1}) and 
(b) the full vertex function $\Gamma_{\mathbf k \mathbf k'}^{\sigma \sigma'}(i\nu_n)$ in Eq.~(\ref{eq:OC-3}).
}
\label{fig:diag}
\end{figure}

\begin{figure}
\centering
\centerline{\includegraphics[width=7.5cm,bb=0 0 500 675]{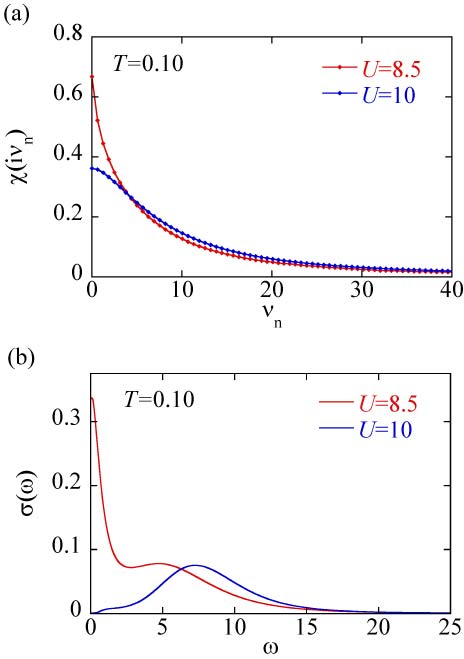}}
\caption{(Color online) (a) Current-current correlation function $\chi(i\nu_n)$ 
and (b) optical conductivity $\sigma(\omega)$ for two $U$'s at $T=0.10$.
}
\label{fig:2GF}
\end{figure}

In order to calculate $\sigma(\omega)$,
we define the current-current correlation functions in Matsubara space as \cite{VC}
\begin{eqnarray}
\chi(i\nu_n)
&=&\sum_{\mathbf k,\sigma}\sum_{i\omega_{l}}v_{\mathbf k}^xv_{\mathbf k}^x\chi_{\mathbf k}^{0,\sigma}(i\nu_n;i\omega_{l})\nonumber \\
&+&\sum_{\mathbf k,\mathbf k',\sigma,\sigma'}\sum_{i\omega_{l},i\omega_{m}'}v_{\mathbf k}^xv_{\mathbf k'}^x\chi_{\mathbf k}^{0,\sigma}(i\nu_n;i\omega_{l})\nonumber \\
&\times& \Gamma_{\mathbf k \mathbf k'}^{\sigma \sigma'}(i\nu_n)\chi_{\mathbf k'}^{0,\sigma'}(i\nu_n;i\omega_{m}'),
\label{eq:OC-1}
\end{eqnarray}
where 
\begin{eqnarray}
\chi_{\mathbf k}^{0,\sigma}(i\nu_n;i\omega_{l})=-\frac{T}{N}
G_{\sigma}(\mathbf k,\omega_{l})G_{\sigma}(\mathbf k,i\omega_{l}+i\nu_n).
\label{eq:OC-8}
\end{eqnarray}
In our calculation as shown in Fig.~\ref{fig:diag} (a), Eq.(\ref{eq:OC-1}) is exact except that the dependence on internal frequencies
in the full vertex function $\Gamma_{\mathbf k \mathbf k'}^{\sigma \sigma'}(i\nu_n;i\omega_{l},i\omega_{m}')$ has been averaged over,
namely, $\Gamma_{\mathbf k \mathbf k'}^{\sigma \sigma'}(i\nu_n;i\omega_{l},i\omega_{m}')\approx \Gamma_{\mathbf k \mathbf k'}^{\sigma \sigma'}(i\nu_n)$.
The preceding studies on the vertex corrections have reported that in the case of spin susceptibility,
this approximation qualitatively reproduced the results consistent with full calculation.\cite{SC-DMFT,SC-DMFT-1,SC-DMFT-2,SC-DMFT-3,SC-DMFT-4}
Since full calculation for the vertex function also requires much longer computational time,
we use this approximation in this study.
This is an uncontrolled approximation and it is very important to check its validity for the current vertex
and this should be studied in future.

To obtain $\Gamma_{\mathbf k \mathbf k'}^{\sigma \sigma'}(i\nu_n)$,
we first calculate two-electron Green's functions in the cluster
$\chi_{\alpha \beta \gamma \delta}^{\sigma \sigma'}(\tau)$$=$$\langle c_{\alpha\sigma}^\dagger(\tau)c_{ \beta\sigma}(\tau)c_{\gamma \sigma'}^\dagger(0) c_{\delta \sigma'}(0)\rangle$
with imaginary time $\tau$ by using CTQMC method, where $\alpha$-$\delta$ denote sites in the cluster.
We then evaluate the irreducible vertex function in the cluster
$I_{\alpha \beta \gamma \delta}^{\sigma \sigma'}(i\nu_n)$ via the Bethe-Salpeter equation,
\begin{eqnarray}
\chi_{\alpha \beta \gamma \delta}^{\sigma \sigma'}(i\nu_n)&=&\chi_{\alpha \beta \gamma \delta}^{0,\sigma \sigma'}(i\nu_n)
+\sum_{\alpha' \beta',\gamma' \delta'}\chi_{\alpha \beta \alpha' \beta'}^{0,\sigma \sigma''}(i\nu_n)\nonumber \\
&\times& I_{\alpha' \beta' \gamma' \delta'}^{\sigma'' \sigma'''}(i\nu_n)\chi_{\gamma' \delta' \gamma \delta}^{\sigma''' \sigma'}(i\nu_n).
\label{eq:OC-2}
\end{eqnarray}
$\chi_{\alpha \beta \gamma \delta}^{0,\sigma \sigma'}(i\nu_n)\equiv -T\sum_{i\epsilon_{l}} G_{\sigma}^{\delta \alpha}(i\epsilon_{l})
G_{\sigma'}^{\beta \gamma}(i\epsilon_{l}+i\omega_n)\delta_{\sigma \sigma'}$.
We then calculate the lattice irreducible vertex
\begin{eqnarray}
I_{\mathbf k,\mathbf k'}^{\sigma \sigma'}(i\nu_n)=\sum_{\alpha \beta \gamma \delta}
I_{\alpha \beta \gamma \delta}^{\sigma \sigma'}(i\nu_n)e^{i\mathbf k \cdot (\mathbf r_{\alpha}-\mathbf r_{\beta})
+i\mathbf k' \cdot (\mathbf r_{\gamma}-\mathbf r_{\delta})},
\label{eq:OC-9}
\end{eqnarray}
and obtain $\Gamma_{\mathbf k \mathbf k'}^{\sigma \sigma'}(i\nu_n)$ via the Bethe-Salpeter equation as shown in Fig.~\ref{fig:diag} (b),
\begin{eqnarray}
\Gamma_{\mathbf k \mathbf k'}^{\sigma \sigma'}(i\nu_n)&=&I_{\mathbf k \mathbf k'}^{\sigma \sigma'}(i\nu_n)
+\sum_{\mathbf k'',\sigma''}\sum_{i\omega_{l}}I_{\mathbf k \mathbf k''}^{\sigma \sigma''}(i\nu_n)\nonumber \\
&\times& \chi_{\mathbf k''}^{0,\sigma''}(i\nu_n;i\omega_{l})
\Gamma_{\mathbf k'' \mathbf k'}^{\sigma'' \sigma'}(i\nu_n).
\label{eq:OC-3}
\end{eqnarray}

Once $\Gamma_{\mathbf k \mathbf k'}^{\sigma \sigma'}(i\nu_n)$ is obtained, we calculate $\sigma(i\nu_n)$ from Eq.~(\ref{eq:OC-1}).
Finally, we calculate real frequency quantity $\sigma(\omega)$ via analytic continuation $i\nu_n\rightarrow \omega+i0 $
by using the maximum entropy method.\cite{MEM}
In what follows, we normalize $\sigma(\omega)$ by the unit of $e^2/\hbar$, where $\hbar$ is the reduced Planck's constant,
and $\omega$ are in units of $t$.

In Fig.~\ref{fig:2GF} (a), we show the results of $\chi(i\nu_n)$ for two $U$'s at $T=0.10$.
As mentioned in the previous subsection, negative sign problem in our MC calculations is not serious,
and then we can calculate $\chi(i\nu_n)$ with high accuracy.
For example, at $U=8.5$, we perform $\sim 10^8$ MC sweeps and averaging over 512 samples and
the relative error $\Delta \chi(i\nu_n)$ is small, e.g. $\Delta \chi(0)/ \chi(0) \sim 0.03$.
By using the maximum entropy method, we obtain $\sigma(\omega)$ from $\chi(i\nu_n)$ as shown in Fig.~\ref{fig:2GF} (b). 
In Sec.~\ref{sec:TP}, we will explain the results in detail.

\section{\label{sec:ThC}Thermodynamic criticality}

In this section, we will examine Mott criticality in thermodynamics by our CDMFT calculations
before discussing electric transport.

\begin{figure}
\centering
\centerline{\includegraphics[width=8cm,bb=0 0 550 1150]{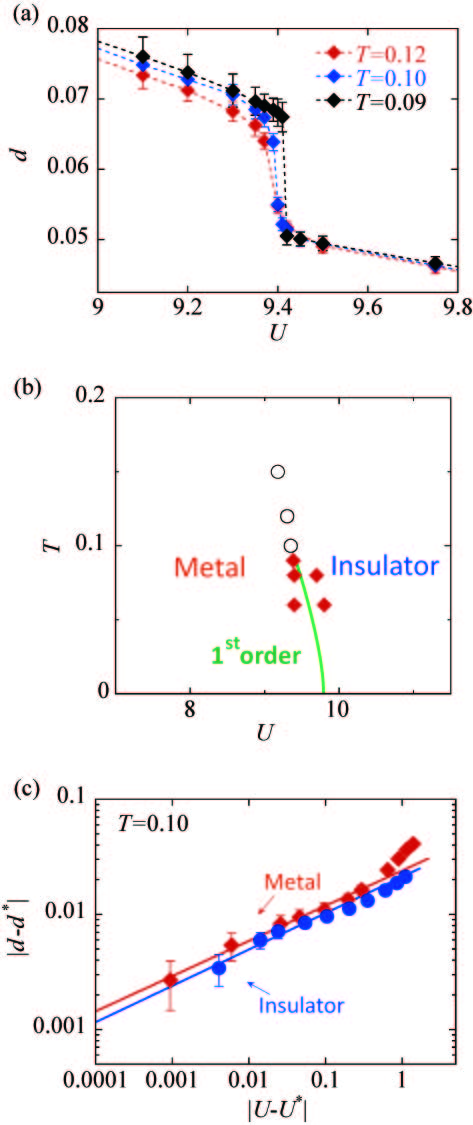}}
\caption{(Color online) (a) Dependence on $U$ of double occupancy $d$ for various $T$'s.
(b) $U$-$T$ phase diagram.
Diamonds represent the boundaries of the region where metallic and insulating solutions coexist.
Line of the first order transition is a guide for eyes.
Circles show the crossover above the Mott critical end point.
(c) Scaling analysis of $d$ upon controlling $U$ at $T=0.10$.
Symbols show calculated data and lines are the results of the fitting Eq.~(\ref{eq:docc-scal}).
}
\label{fig:3-1}
\end{figure}

We first start with identifying the location of the finite-$T$ Mott transition and determine $U$-$T$ phase diagram.\cite{phase}
To this end, we investigate the singularity of the double occupancy 
\begin{eqnarray}
d=\langle n_{i\uparrow}n_{i\downarrow}\rangle
\label{eq:docc}
\end{eqnarray}
with varying $U$ at several $T$'s.
Figure.~\ref{fig:3-1}~(a) presents $U$-dependence of $d$ for various $T$'s.
At lower temperature $T=0.09$, $d$ shows a jump and hysteresis, corresponding to the first-order Mott transition.
At higher temperature $T=0.12$, $d$ changes smoothly and there is no hysteresis in the $U$-dependence.
This indicates a crossover between metal and insulator.
We have checked that the jump and hysteresis of $d$ shrink with increasing $T$ and disappear for $T \geq 0.10$.
Repeating calculations for various $T$'s, the Mott critical end point is found to be at $U^* \cong 9.4$ and $T^* \cong 0.10$.
We determine the $U$-$T$ phase diagram and the result is shown in Fig.~\ref{fig:3-1}~(b).
In the following, we will analyze a singularity of $d$ around the Mott critical end point.

In the Mott transition, $d$ plays the role of order parameter.\cite{docc-op}
Previous single-site DMFT and four-site CDMFT calculations show that $U$-dependence of $d$ is well fitted by a power-law scaling function
with the mean filed exponent $1/\delta_{\rm MF}=1/3$ near the Mott critical end point.\cite{MTC-dcc-DMFT-1,docc-scaling}
This confirms that the Mott transition belongs to Ising universality class.
These approaches are still a mean field approximation, and therefore the obtained critical exponent is the mean field value.
Let us examine our results of $d$ at $T=0.10$.
$d$ changes continuously with $U$ but exhibits a divergent slope near $U=9.4$.
We try to fit the curve with a power-law scaling function,
\begin{eqnarray}
|d-d^*|=A^{\pm}|U-U^*|^{1/\delta^{\pm}},
\label{eq:docc-scal}
\end{eqnarray}
where the subscript $-(+)$ denotes the metallic (insulating) side.
Figure~\ref{fig:3-1} (c) shows the result of scaling analysis for $d$ using five points on each side
around the critical point.
This shows that the fitting is successful on both sides, namely, $d$ exhibits the critical behavior.
The obtained critical exponents are
\begin{eqnarray}
(1/\delta^{-}, 1/\delta^{+})\sim(0.32 \pm 0.05, 0.30 \pm 0.04),
\label{eq:docc-exp}
\end{eqnarray}
and our CDMFT calculations also reproduce the mean field exponent within accuracy.
Here, we estimated $1/\delta^{\pm}$ and those standard deviation by repeating scaling analysis
for each of the 64 data sets of $d$.

\begin{figure}
\centering
\centerline{\includegraphics[width=7.5cm,bb=0 0 500 350]{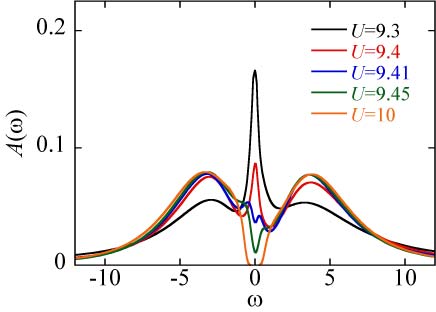}}
\vspace{-5mm}
\caption{(Color online) Density of states $A(\omega)$ for various $U$'s at $T=0.10$.}
\label{fig:dos}
\end{figure}

\section{\label{sec:EP}Electronic Properties near the Mott critical end point}

In the previous section, we have examined the finite-$T$ Mott transition and obtained the $U$-$T$ phase diagram as shown in Fig.~\ref{fig:3-1}~(b).
Next, we will study electronic properties, particularly near the Mott critical end point.
This is useful for understanding electric transport, which will be discussed in Sec.~\ref{sec:TP}.

Density of states is the most fundamental quantity of electronic structure and it has been calculated by
several groups for the triangular lattice.\cite{OC-DMFT-2,A-1,A-2}
These results demonstrated the existence of a quasiparticle peak near the Fermi level in the metallic region and
its disappearance at the metal-insulator transition resulting into the formation of a gap at the Fermi level.
However, detailed behaviors near the Mott critical end point are not well understood, and we examine them.

Density of states $A(\omega)$ is calculated from the lattice Green's function in Eq.~(\ref{eq:lattice-3}) as
\begin{eqnarray}
A(\omega)=-\frac{1}{N\pi}\sum_{\mathbf k}{\rm Im}G(\mathbf k, \omega+i0).
\label{eq:DOS}
\end{eqnarray}
Here, the Green's function $G_{\sigma}(\mathbf k, \omega)$ does not depend on spin $\sigma$ since we treat only paramagnetic solution in CDMFT calculations
and we drop the spin index.
To obtain $A(\omega)$, we have carried out analytic continuation $i\omega_n\rightarrow \omega+i0 $ by using the maximum entropy method.~\cite{MEM}

Figure~\ref{fig:dos} presents the change of $A(\omega)$ with varying $U$ at $T=0.10$ fixed.
At all the values of $U$, $A(\omega)$ shows broad peaks at $U \sim \pm 4$, and they correspond to the upper and lower Hubbard bands.
At $U=9.3$, in addition to them, $A(\omega)$ shows a sharp peak at $\omega=0$.
This is a quasiparticle peak: indication of the metallic state.
At larger $U=10$, the quasiparticle peak disappears and there emerges a gap around $\omega=0$
and its existence is a characteristics of the insulating state.

The interesting characteristic is that two small peaks appear around $\omega=0$ on the insulating side near the Mott transition, for example, at $U=9.41$.
With approaching the Mott transition, their peaks shift towards $\omega=0$ and grow their intensity.
These peaks continue to the quasiparticle peak on the metallic side.
We have also confirmed this behavior for various $T$'s, and have found that the emergence of two small peaks is clearer near the Mott critical end point.
This is more complicated than the canonical picture of Mott transition.
The canonical picture is that a quasiparticle peak disappears at the metal-to-insulator transition and a gap opens continuously.
Our calculations show that this picture is supplemented by the appearance of the low-energy small peaks on the insulating side near the Mott transition.

\begin{figure}
\centering
\centerline{\includegraphics[width=7cm,bb=0 0 500 1200]{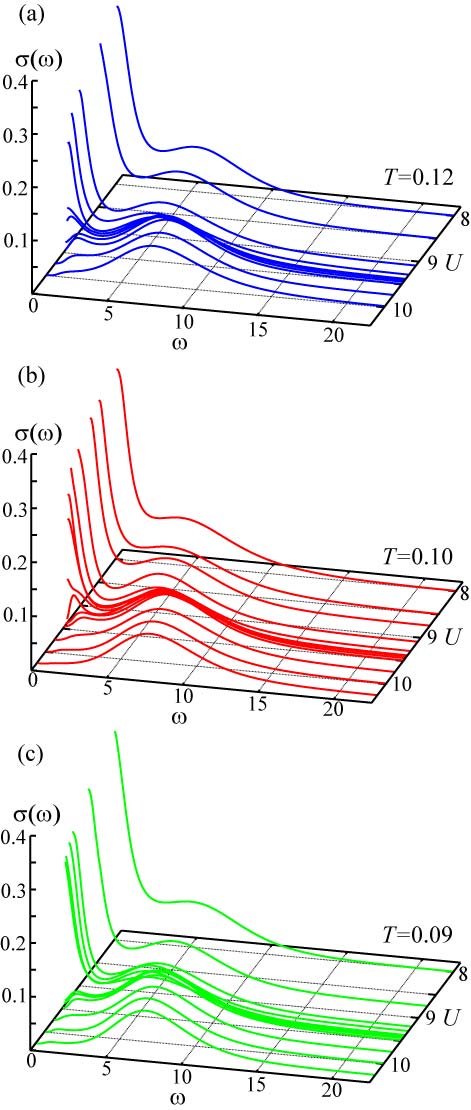}}
\caption{(Color online) Optical conductivity $\sigma(\omega)$ for various $U$'s
at (a) $T=0.12$, (b) $T=0.10$, and (c) $T=0.09$.
}  
\label{fig:5-1-2}
\end{figure}

\section{\label{sec:TP}Transport Properties}

In this section, we will investigate optical conductivity $\sigma(\omega)$.

First, let us summarize the general trends of $\sigma(\omega)$ in the $U$-$T$ space.
Figures~\ref{fig:5-1-2} (a)-(c) show the variation of $\sigma(\omega)$ upon controlling $U$ at several $T$'s.
First, metallic state is realized for small $U$ at all $T$'s.
$\sigma(\omega)$ shows a Drude peak at low $\omega$ as expected, indicating the formation of coherent quasiparticle.
There exists a broad peak at $\omega \sim U$ and this comes from excitations to the Hubbard band.
Next, insulating state is realized for large $U$ at all $T$'s.
The Drude peak disappears and its absence is a characteristic of insulator.
The most important part is near the Mott critical end point $T=0.10$ and $U \sim 9.4$ and we investigate $\sigma(\omega)$ there
upon controlling $U$ at various $T$'s.
At $T=0.10$, we find that $\sigma(\omega)$ at low $\omega$ changes quickly at $U \sim 9.4$, 
At the lower temperature $T=0.09$, $\sigma(\omega)$ at low $\omega$ shows a jump, corresponding to the first-order transition,
and we also confirm that there is a hysteresis. 
At the higher temperature $T=0.12$, $\sigma(\omega)$ at low $\omega$ shows a smooth crossover from metallic to insulating state.
In the following, we will analyze $\sigma(\omega)$ on metallic and insulating sides in detail,
particularly, focusing on its dependence on $U$ at $T=0.10$ near the Mott critical end point.

\begin{figure}
\centering
\centerline{\includegraphics[width=7.5cm,bb=0 0 500 550]{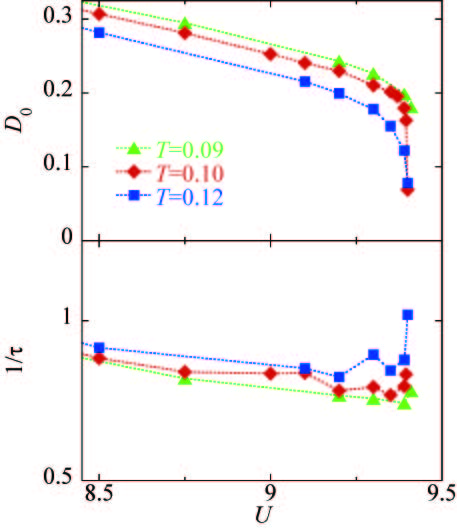}}
\caption{(Color online) Dependence on $U$ of a Drude weight $D_{\rm 0}$ (upper panel) and a transport scattering rate $1/\tau$ (lower panel)
at various $T$'s.
}  
\label{fig:5-3}
\end{figure}

\subsection{\label{sec:level1}Metallic side}

For simple metals at low temperature, it is well known that the low-$\omega$ part of $\sigma(\omega)$
is described by a simple Drude formula
\begin{eqnarray}
\sigma(\omega)={\rm Re}\Biggl[\frac{D_{\rm 0}}{-i\omega+\frac{1}{\tau}}\Biggr],
\label{eq:Drude}
\end{eqnarray}
%
where $D_{\rm 0}$ is a Drude weight and $1/\tau$ is the transport scattering rate.
We fit the low-$\omega$ peak of our data by Eq.~(\ref{eq:Drude}) and examine the changes in $D_{\rm 0}$ and $1/\tau$ upon controlling $U$ at $T$ fixed.

Figures~\ref{fig:5-3} (a) and (b) show $U$-dependence of $D_{\rm 0}$ and $1/\tau$ for various $T$'s.
One may expect that with approaching the Mott critical end point, $1/\tau$ increases with increasing $U$ and diverge.
One can see that $1/\tau$ does not change noticeably with $U$ and shows small increase near the Mott transition point.
$D_{\rm 0}$ shows noticeable decrease with $U$ and drops drastically near the Mott transition point at $T=0.10$.
These results suggest that Mott transition is driven by the change in $D_{\rm 0}$ rather than $1/\tau$.

\subsection{\label{sec:level1}Insulating side}

\begin{figure}
\centering
\centerline{\includegraphics[width=7.5cm,bb=0 0 500 1000]{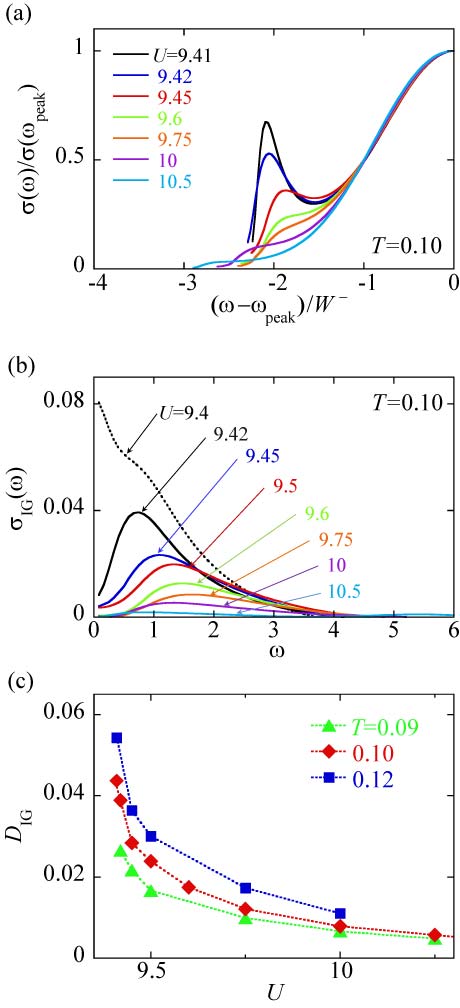}}
\caption{(Color online) (a) Fitting results of an incherent broad peak on the insulating side
by Eq.~(\ref{eq:HBfit-1}).
(b) Ingap peak $\sigma_{\rm IG}(\omega)$ on the insulating side.
At $U=9.4$, a small Drude peak appears, accompanied by an ingap peak around $\omega \sim 1$.
(c) Dependence on $U$ of the weight of an ingap peak $D_{\rm IG}$ for various $T$'s.
}
\label{fig:5-4}
\end{figure}

\begin{figure}
\centering
\centerline{\includegraphics[width=6cm,bb=0 0 450 1000]{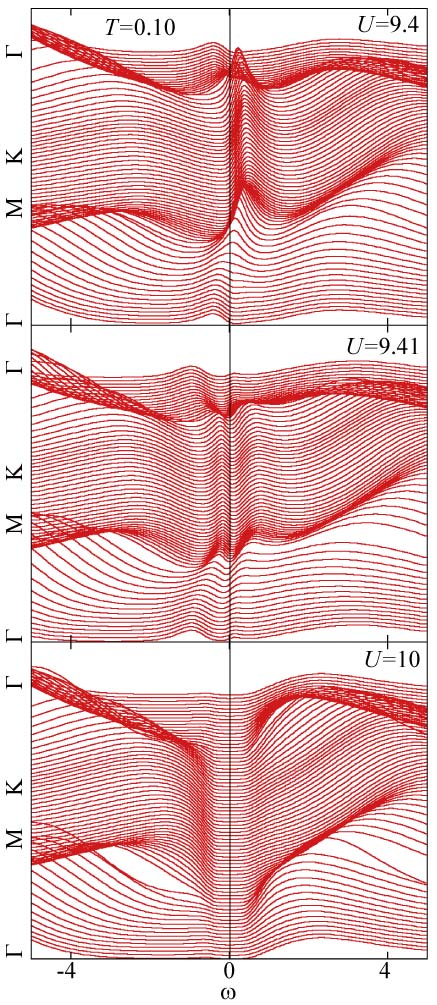}}
\caption{(Color online) $\mathbf k$-resolved spectral function $A_{\mathbf k}(\omega)$  at three values of $U$ at $T=0.10$
along the path $\Gamma$-$\rm K$-$\rm M$-$\Gamma$ in Fig.~\ref{fig:CDMFT}~(c).}
\label{fig:5-8}
\end{figure}

Now, let us study the insulating side near the Mott transition.
The most important difference from conventional insulators is that $\sigma(\omega)$ has two parts;
one is a broad Hubbard peak around $\omega \sim U$ and the other is a small peak at low $\omega$.
The former peak persists from the metallic side and comes from the contribution of the Hubbard band, $\sigma_{\rm HB}(\omega)$.
Its peak position shifts from the strong coupling limit $\omega \sim U$ by a finite amount $\sim -3$,
which comes from the kinetic energy of a pair of doublon and holon.
We find that $\sigma_{\rm HB}(\omega)$ is well fitted by a simple Gaussian form,
\begin{eqnarray}
\frac{\sigma_{\rm HB}(\omega)}{\sigma(\omega_{\rm peak})}
=A \rm exp \Bigl[\frac{\it -B(\omega-\omega_{\rm peak})^2}{\it W_{-}^{\rm 2}}\Bigr],
\label{eq:HBfit-1}
\end{eqnarray}
particularly for $\omega$ below the peak position $\omega_{\rm peak}$ of $\sigma_{\rm HB}(\omega)$.
This is shown in Fig.~\ref{fig:5-4}~(a).
Here, $A$ and $B$ are constant and $W_{-}$ is the half width below $\omega_{\rm peak}$.

In order to carry out detailed analysis of the small peak at low $\omega$,
we subtract $\sigma_{\rm HB}(\omega)$ from the total $\sigma(\omega)$ by Eq.~(\ref{eq:HBfit-1}) and the remaining part is defined as $\sigma_{\rm IG}(\omega)$.
The results are shown in Fig.~\ref{fig:5-4}~(b).
They all show a peak below $\omega \sim 2$.
The peak position shifts toward lower $\omega$ with approaching the Mott transition point.
We will call this structure an $\it ingap$  peak hereafter.
For this part, we define its intensity by,
\begin{eqnarray}
D_{\rm IG}=\frac{2}{\pi} \int_{0}^{\infty} d\omega \sigma_{\rm IG}(\omega).
\label{eq:DIG-1}
\end{eqnarray}
We analyze our data in the insulating phase and show the $U$-dependence of $D_{\rm IG}$ in Fig.~\ref{fig:5-4}~(c).
$D_{\rm IG}$ increases with decreasing $U$ and its growth is drastic at $T=0.10$
near the Mott critical end point.

\section{\label{sec:IG}Ingap peak and spectral function}

We have shown that optical conductivity $\sigma(\omega)$ exhibits an ingap peak, an interesting structure
on the insulating side near the Mott transition.
In this section, we briefly discuss the ingap peak from a viewpoint of electronic structure.

In Sec.~\ref{sec:EP}, we examined the density of states $A(\omega)$ and found that $A(\omega)$ exhibits low-energy two small peaks
upon approaching the Mott transition point from the insulating side.
To see its origin, let us calculate the $\mathbf k$-resolved spectral function $A_{\mathbf k}(\omega)$.
Two previous studies also calculated $A_{\mathbf k}(\omega)$ in a triangular lattice and reported two features
with compared to the square lattice case.
One is weaker intensity of quasiparticle peaks in the metallic phase, and the other is the absence of anomolous
pseudogap phase.\cite{A-1,A-2}
Ohashi~{\it et al.} studied an anisotropic triangular lattice system and observed the split of the quasiparticle peak
in the insulating phase, which is attributed to enhanced magnetic fluctuations.\cite{Ak-1}

Figure~\ref{fig:5-8} presents $A_{\mathbf k}(\omega)$ at three values of $U$ at $T=0.10$.
The data are plotted along the path $\Gamma$-$\rm K$-$\rm M$-$\Gamma$ in the Brillouin zone shown in Fig.~\ref{fig:CDMFT}~(c).
At the weakest correlation $U=9.4$, in addition to broad peaks corresponding to the upper and lower Hubbard bands,
a quasiparticle peak exists around $\omega=0$.
The energy dispersion is strongly renormalized to a smaller width due to strong correlation effects.
We have confirmed that the Fermi surface is nearly spherical.
The quasiparticle peak disappears at the stronger correlation $U=9.41$ and there emerges two small peaks
centered around $\omega=0$, which exist in the whole Brillouin zone.
At the strongest correlation $U=10$, these small peaks disappear and there emerges a clear gap around $\omega=0$.

We have found that the small peaks start to appear at the same time of emergence of the ingap peak in $\sigma(\omega)$.
We have also checked that the energy difference between the two small peaks agrees with the position of the ingap peak. 
Therefore, the ingap peak is an optical transition between these two peaks, which is associated with the collapse of quasiparticles.

\section{\label{sec:TrC}Transport Criticality}

In this section, we provide a detailed analysis of optical conductivity $\sigma(\omega)$ near the Mott critical end point to 
discuss the Mott criticality in electric transport.

We first examine whether $\sigma(\omega)$ exhibits a singularity at the Mott critical end point.
The most important issue is how to identify a singularity of $\sigma(\omega)$, and this has not been well established.
Since $\sigma(\omega)$ is not derived from free energy, protocol of scaling analysis is not uniquely determined 
like that for thermodynamic quantities.
Therefore, we choose several quantities in $\sigma(\omega)$ as a scaling variable and examine their criticality.

\begin{figure}
\centering
\centerline{\includegraphics[width=7.5cm,bb=0 0 500 700]{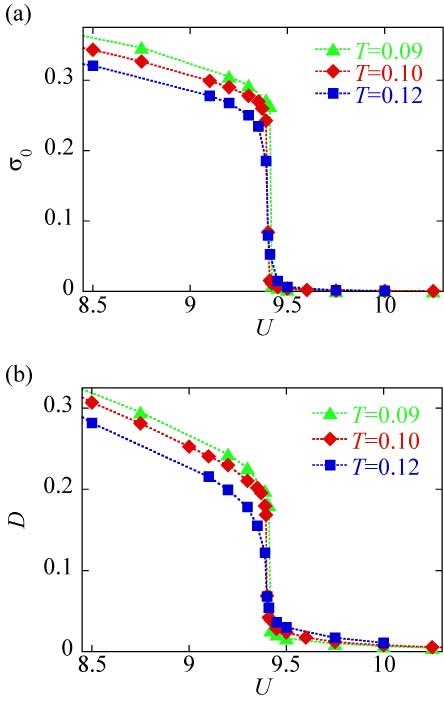}}
\caption{(Color online) Dependence on $U$ of (a) dc-electric conductivity $\sigma_{0}$ and optical weight $D$ for various $T$'s.}
\label{fig:5-5}
\end{figure}

The simplest choice as a scaling variable is dc-electric conductivity $\sigma_{0}=\sigma(\omega=0)$.
This has been used in experimental studies on the Mott criticality in electric transport.~\cite{MTC-exp-V,MTC-exp-kapp}
The experimental results have demonstrated the existence of a singularity of $\sigma_{0}$ near the Mott critical end point.
However, $\sigma_{0}$ is not necessarily the best choice because impurity scatterings may contaminate its intrinsic critical behavior.
In fact, scaling analysis of the experimental data has been performed only on the metallic side.
Here, our results of $U$-dependence of $\sigma_{0}$ at various $T$'s are plotted in Fig.~\ref{fig:5-5}~(a).
At $T=0.10$, $\sigma_{0}$ shows a singularity near $U=9.4$.
However, its scaling analysis on the insulating side leads to rather large relative errors, since $\sigma_{0}$ is very small, as we will explain later.

\begin{figure}
\centering
\centerline{\includegraphics[width=8cm,bb=0 0 500 350]{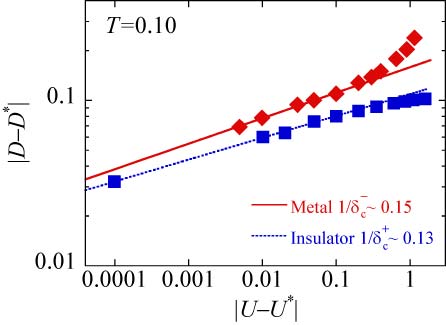}}
\caption{(Color online) Scaling analysis of optical weight $D$.
Symbols show calculated data and lines are the results of the fitting Eq.~(\ref{eq:scal-1}).}  
\label{fig:5-7}
\end{figure}

An alternative scaling variable may be the optical weight of a Drude peak $D_{\rm 0}$ on the metallic side.
This quantity is obtained by integrating the low-$\omega$ peak of $\sigma(\omega)$, and then is more robust against impurity scatterings.
However, the insulating side does not have the Drude peak.
In Sec.~\ref{sec:TP}, we have shown that with increasing $U$, the Drude peak on the metallic side changes to the ingap peak on the insulating side.
Therefore, a candidate on the insulating side is the optical weight of an ingap peak $D_{\rm IG}$.
We show the $U$-dependence of the total optical weight $D=D_{\rm 0}+D_{\rm IG}$ for various $T$'s in Fig.~\ref{fig:5-5}~(b).
At higher temperature $T=0.12$, $D$ shows a smooth crossover from the metallic to the insulating side.
At lower temperature $T=0.09$, $D$ shows a jump near $U=9.4$. 
We have also checked the existence of hysteresis, and these results indicate the first-order Mott transition.
At $T=0.10$, $D$ changes drastically near $U=9.4$ without showing its hysteresis similar to $\sigma_{0}$.

In the following, we will examine a singularity of $D$ in detail, particularly its dependence on $U$ at $T=0.10$.

\begin{figure}
\centering
\centerline{\includegraphics[width=7.5cm,bb=0 0 500 550]{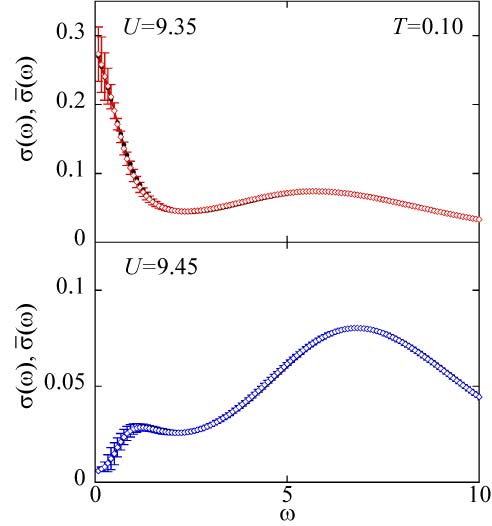}}
\caption{(Color online) 
Optical conductivity for two $U$'s at $T=0.10$; $\sigma(\omega)$ ($\blacklozenge$) and $\bar{\sigma}(\omega)$ ($\lozenge$).
Error bars are estimated by the jackknife analysis.
}  
\label{fig:5-6-1}
\end{figure}

\subsection{\label{sec:SOW}Preliminary scaling analysis and accuracy of optical conductivity}

Let us start a scaling analysis of the optical weight at $T=0.10$.
To perform more elaborate scaling analysis, we estimated error in optical conductivity.
In this subsection, we show these results and proceed to the scaling analysis of its optical conductivity in the next subsection.

As mentioned in Sec.~\ref{sec:ThC}, criticality in thermodynamics has been investigated by examining the thermal average double occupancy $d$
as a function of $T$ or $U$.
Following this analysis, we try to fit the curve of the optical weight $D$ with a power-law scaling function, 
\begin{eqnarray}
|D-D^*|=A^{\pm}|U-U^*|^{1/\delta_{c}^{\pm}},
\label{eq:scal-1}
\end{eqnarray}
where the subscript $-(+)$ denotes the metallic (insulating) side.
$D^*$ denotes the value at the Mott critical end point $(T^*, U^*)$ and $A^{\pm}$ is a constant.
The critical exponents $1/\delta_{c}^{\pm}$ reflect the underlying universality class.
Fitting parameters are $1/\delta_{c}^{\pm}$, $A_{\pm}$, $D^*$ and $U^*$.
Figure~\ref{fig:5-7} shows this result of scaling analysis for Eq.(\ref{eq:scal-1}) using five points on each side of the Mott critical end point.
The fitting is successful on both sides, and $D$ exhibits a critical power-law behavior.
The critical region is about $|U-U^*|\leq 0.2$, which is similar to that of $d$ discussed in Sec.~\ref{sec:ThC}. 
In this scaling analysis, the critical end point is determined simultaneously and $(U^*, D^*)\sim(9.40, 0.11)$.
The obtained critical exponents are $(1/\delta_{c}^{-}, 1/\delta_{c}^{+})\sim(0.15, 0.13)$.

This is a preliminary scaling and we need error analysis of $D$ for more reliable value of the critical exponent.
In the maximum entropy method, the error in imaginary-time QMC samplings does not transfer to the calculating real-frequency data.
Here, we use the jackknife method for an error analysis of optical conductivity.\cite{JaNa-1}
In the jackknife method, we divide total $N$ sets of data of a quantity $O$ into $N_b$ bins of the bin size $m=N/N_{b}$.
For each bin $b$, we first define the average $O_b$ by taking away bin $b$ from the whole data.
For these $N_b$ data, we then define their average $\bar{O}$ and error $\Delta \bar{O}$,
\begin{eqnarray}
\bar{O}&=&\frac{1}{N_b}\sum_{b=1}^{N_b}O_b,\nonumber\\
\Delta \bar{O}&=&\biggl[\frac{N_b-1}{N_b}\sum_{b=1}^{N_b}\bigl(O_b-\bar{O} \bigr)^2\biggr]^{\frac{1}{2}}.
\label{eq:JN-1}
\end{eqnarray}
One checks the dependence of $\Delta \bar{O}$ on $m$, and if they are nearly constant, one can use that value 
as the error of $O$.
For optical conductivity, we have performed this analysis with varying $m$ from 2 to 64 for $N=512$ data sets.
We have checked the dependence of $\Delta \bar{O}$ on $m$, and found that the errors are almost constant
for all $\omega$'s over the whole range of bin sizes.
In the rest of the paper, we will show the average and the error of optical conductivity and other quantities
that are analyzed with the bin size $m=16$ unless explicitly mentioned.

Figure~\ref{fig:5-6-1} presents the results of optical conductivity $\bar{\sigma}(\omega)$ at $T=0.10$
on the metallic ($U=9.35$) and insulating ($U=9.45$) sides.
We find that $\bar{\sigma}(\omega)$ agrees with $\sigma(\omega)$
within accuracy in both cases.
The relative error is small at high $\omega$'s, e.g. $\Delta \bar{\sigma}/\bar{\sigma} \sim 0.01$
at $\omega \sim U=9.35$ and $9.45$, while the error at small $\omega$'s is larger, e.g. $\Delta \bar{\sigma}/\bar{\sigma} \sim 0.12$ and $0.22$
at $\omega \sim 0$ for $U=9.35$ and $9.45$, respectively.
Having the error in optical conductivity, we proceed to estimating the error in optical weight and its scaling analysis.

\begin{figure}
\centering
\centerline{\includegraphics[width=8cm,bb=0 0 600 350]{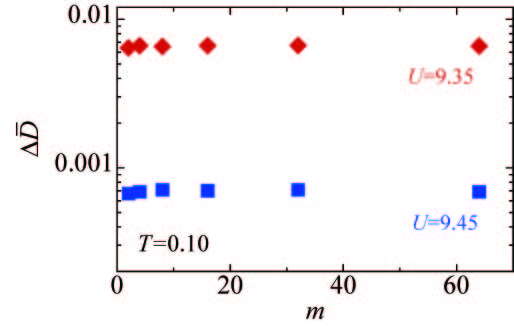}}
\caption{(Color online) 
Bin size $m$-dependence of jackknife error in optical weight $\Delta \bar{D}$.
}  
\label{fig:JN}
\end{figure}

\begin{figure}
\centering
\centerline{\includegraphics[width=8.1cm,bb=0 0 550 1200]{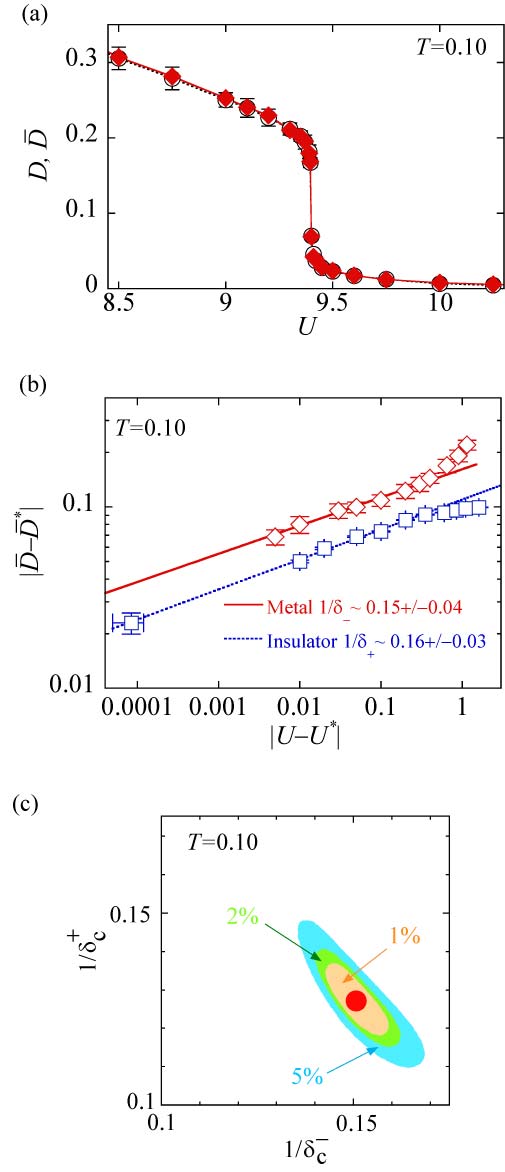}}
\caption{(Color online) 
(a) $U$-dependence of optical weight; $D$ ($\blacklozenge$) and $\bar{D}$ ($\bigcirc$).
(b) Scaling analysis of $\bar{D}$ when errors are evaluated.
Open symbols show calculated data, dotted lines are the fitting results and error bars are estimated by the jackknife analysis.
(c) Error in the critical exponents $1/\delta_{\pm}$ due to fitting.
}  
\label{fig:5-6-2}
\end{figure}

\begin{figure}
\centering
\centerline{\includegraphics[width=8cm,bb=0 0 500 600]{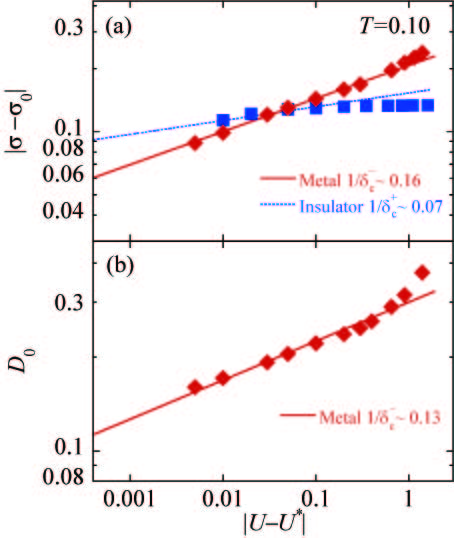}}
\caption{(Color online) Scaling analysis of (a) dc-electric conductivity $\sigma_{0}$ and (b) Drude weight $D_{\rm 0}$.
Symbols show calculated data and lines are the fitting results.
}  
\label{fig:dc-cond}
\end{figure}

\subsection{\label{sec:SOW}Critical exponent of optical weight}

In the previous subsection, we have estimated the error in optical conductivity.
We will evaluate the error in the critical exponents of optical weight, and proceed to their discussion in detail.

From the $N=512$ sets of optical conductivity, we calculate the optical weight for all and then determine
their average $\bar{D}$ and error $ \Delta \bar{D}$  by the jackknife method.
The bin size dependence of  $ \Delta \bar{D}$ is shown in Fig.~\ref{fig:JN} for two $U$'s.
This dependence is nearly constant for this and all the other values of $U$, and this shows that
the jackknife error analysis works well in our calculations.
It is important that the relative error $\Delta \bar{D}/\bar{D}$ is much smaller than $\Delta\bar{\sigma}/\bar{\sigma}$, e.g. 
$\Delta \bar{D}/\bar{D} \sim 0.03$ for $U=9.35$ and $9.42$.
The average $\bar{D}$ agrees with $D$ calculated at the beginning of Sec.\ref{sec:TrC} within the error $ \Delta \bar{D}$ in Fig.~\ref{fig:5-6-2} (a).
Note that $D$ is an average defined differently.
For this, the average is taken over the imaginary-time data, and this averaged result is analytically continued to real-frequency and
finally the optical weight $D$ is calculated.

Now, with these data of $\bar{D}$, let us perform the scaling analysis.
For each of the $N_b$ data of the optical weight, we repeat the scaling analysis Eq.(\ref{eq:scal-1}),
and estimate the critical exponents $1/\delta_{c}^{\pm}$ by the jackknife method.
Their average and error are
\begin{eqnarray}
(1/\delta_{c}^{-}, 1/\delta_{c}^{+})\sim(0.15\pm0.04, 0.16\pm0.03).\nonumber
\label{eq:exp-1}
\end{eqnarray}
Fig.~\ref{fig:5-6-2} (b) shows the fitting with these values,
and this shows that the fitting is successful on both of the metallic and insulating sides.

Finally, we discuss the value of $1/\delta_{c}^{\pm}$.
In Sec.\ref{sec:ThC}, we have examined the critical behavior of double occupancy and have confirmed the mean-field value of Ising
critical exponent $1/\delta=1/3$. 
If conductivity has the same critical behaviors as claimed by Limelette~{\it et al.},\cite{MTC-exp-V}
the optical weight should have the same critical exponent and therefore $1/\delta_{c}^{\pm}=1/3$ in our case.
This is because these two are calculated in the same approximation, i.e. CDMFT.
However, our result of the critical exponent is not close to $1/3$ or any of these values in the Ising universality class
($1/15$ and $1/4.8$ for the two- and three-dimensional cases~\cite{LC-tri-1}).

We have also estimated the error in the critical exponents due to fitting.
Figure~\ref{fig:5-6-2} (c) shows the accuracy of our fitting.
The residue of fitting is plotted in the space of the two fitting parameters $1/\delta_{c}^{\pm}$.
The order parameters are all optimized for each pair of $1/\delta_{c}^{\pm}$.
The residue is shown by its relative difference compared to the minimum value: i.e. 5\%
means that the residue is larger by 5\% than that of the globally optimized set of the fitting parameters.
This plot shows that the exponent of double occupancy $1/\delta=1/3$ has a very large residue
and this value is very unlikely.
Thus we can exclude the possibility that the optical weight has the same critical behavior as the order parameter
in the Mott transition.

Recently, Papanikolaou~{\it et al.} suggested a different scenario for the scaling law of conductivity;
the subleading term due to energy density dominates rather than the leading term (magnetization).\cite{MTC-kapptheo-2}
If that is the case, one may see that the mean-field critical exponent of energy density $1/\delta_{\rm E}=1/1.5$.
However, the obtained critical exponent is not close to $1/\delta_{\rm E}$ either.

We have also tried the same analysis of other quantities of $\sigma(\omega)$.
Figure~\ref{fig:dc-cond} (a) shows the result for dc-electric conductivity $\sigma_{0}$.
We confirm that $\sigma_{0}$ shows a critical power-law behavior on each of the metallic and insulating sides.
However, the obtained two critical exponents are clearly distinct, $(1/\delta_{c}^{-} \sim 0.16)\ll(1/\delta_{c}^{+} \sim 0.07)$,
while $1/\delta_{c}^{-}$ is close to the one in the optical weight.
We also analyze a Drude weight $D_{0}$ on the metallic side only by the power-law scaling function of $U$,
\begin{eqnarray}
D_{\rm 0}=A^{-}(U^*-U)^{1/\delta_{c}^{-}}.
\label{eq:scal-2}
\end{eqnarray}
As shown in Fig.~\ref{fig:dc-cond} (b), $D_{0}$ shows a critical power-law behavior
and leads to a similar value $1/\delta_{c}^{-} \sim 0.13$ to that in the optical weight.

\section{\label{sec:SD}Summary and discussion}

In this paper, we have studied the Mott criticality in electric transport of the half-filled Hubbard model on a triangular lattice.
We have succeeded in the numerical calculations of optical conductivity taking into account correlation effects and
frustrated lattice geometry by the cellular dynamic mean field theory including vertex corrections. 
%

Before calculating conductivity, we checked thermodynamic properties and determined the phase diagram
and the Mott critical end point within our version of numerical method.
We then examined the criticality of double occupancy, which is the order parameter in Mott transition.
To discuss transport properties, we also investigated electronic structure.
In addition to density of states, which is a local quantity, we calculated single-electron spectra at each wave vector.

The phase diagram determined in our work is consistent with the previous works, and the position of the critical end point is confirmed.
Our calculation also reproduced the expected critical exponent of double occupancy, and that is the mean-field value of the Ising order parameter
with respect to its dual field $1/\delta_{\rm MF}=1/3$.

The density of states shows two small peaks around zero frequency on the insulating side near the Mott transition,
which originate from a quasiparticle peak on the metallic side.
This suggests that the canonical picture of Mott transition is supplemented by the appearance of these small peaks.

We have extensively analyzed optical conductivity and the criticality  of its characteristic quantities.
In this paper, we mainly focused on the criticality upon controlling Coulomb interaction at the critical temperature.
Before performing detailed analysis of transport criticality, we have tried a Drude analysis of optical conductivity.
It indicates that Mott transition is primarily determined by the change in the Drude weight rather than transport scattering rate.
Another important point is that optical conductivity on the insulating side has a small peak at low frequencies, which we named an ``ingap" peak.
The wave-vector resolved single-electron spectra show the ingap peak,  which originates in transitions between two low-energy small peaks around zero frequency
appearing after the collapse of quasiparticles.

The Drude peak on the metallic side continues to the ingap peak on the insulating side with Coulomb repulsion.
We have analyzed the singularity of the optical weight of these peaks near the Mott critical end point 
and have compared its critical exponent with that of double occupancy within the same calculations.
However, the obtained critical exponent of the optical weight differs from that of double occupancy
or corresponding value of Ising exponent in any dimension.

We have justified the belief that double occupancy has the same scaling behavior as the Ising order parameter.
The most important point is that the critical exponent in conductivity is unconventional in contract to that of double occupancy
within our numerical accuracy.
This discrepancy may disprove the appealing and popular working hypothesis that conductivity exhibits the same scaling behavior as order parameter
in the Mott transition.
Regarding another possibility that conductivity corresponds to energy density of the Ising universality class,\cite{MTC-kapptheo-2}
the corresponding mean field exponent differs from our result too. 

In the cellular dynamical mean field calculations, the cluster size is one of the important conditions.
As for double occupancy, preceding studies with single- and four-site clusters both concluded
the same critical exponent as our calculations with a three-site cluster.\cite{MTC-dcc-DMFT-1,docc-scaling}
Therefore, one may expect that the critical exponent in conductivity is also robust against the change of the cluster size,
but we will need to check this point directly using different cluster sizes and geometries.

Another point is about model Hamiltonian.
Our results are obtained for the Hubbard model, i.e., the standard model of a strongly correlated electronic system.
However, some of the terms neglected in the Hubbard model may become important and crucial near a metal-insulator transition.
They include electron-phonon interactions and long-range part of Coulomb interaction.
If they dominate criticality of the transition, the observed exponents differ from those of the Hubbard model.
Our understanding of these models is very limited, and this is a future problem.

Lastly, we make a comment on scaling analysis in experimental works.
Our scaling analysis of dc-electric conductivity shows that two critical exponents on the metallic and insulating sides do not agree.
In the experimental studies,\cite{MTC-exp-V,MTC-exp-kapp} a singularity of dc-electric conductivity has been analyzed
only on the metallic side.
It is interesting to check whether both critical exponents are identical in experiments.
In optical conductivity, it is expected that a low-energy peak appears on the insulating side like an ingap peak.
It is also interesting to observe this peak and examine if it evolves into the Drude peak with varying pressure.

\section*{Acknowledgments}
The authors are grateful to T. Ohashi and H. Kusunose for the advice about Monte Carlo calculations. 
The present work is supported by MEXT Grant-in-Aid for Scientific Research on Priority Areas ``Novel States of Matter Induced by Frustration" (No.~19052003),
and by Next Generation Supercomputing Project, Nanoscience Program, MEXT, Japan.
Numerical computation was performed with facilities at Supercomputer Center in ISSP and Information Technology Center,
University of Tokyo.


\begin{thebibliography}{}
\bibitem{MT-exp-1}D. B. McWhan {\it et al}., Phys. Rev. Lett. {\bf 27}, 941 (1971). 
\bibitem{MI-tri-1}K. Kanoda, J. Phys. Soc. Jpn. {\bf 75}, 051007 (2006).
\bibitem{LC-tri-1}H. E. Stanley, {\it Introduction to phase transition and critical phenomena} (Oxford University Press, 1971).
\bibitem{docc-op}C. Castellani, C. Di Castro, D. Feinberg, and J. Ranninger, Phys. Rev. Lett. {\bf 43}, 1957 (1979).
\bibitem{MI-crit-1}M. Imada, Phys. Rev. B {\bf 72}, 075113 (2005).
\bibitem{MTC-dcc-DMFT-1}G. Kotliar, E. Lange, and M. J. Rozenberg, Phys. Rev. Lett. {\bf 84}, 5180 (2000).
\bibitem{docc-scaling} P. S$\acute{\rm e}$mon and A.-M. S. Tremblay, Phys. Rev. B {\bf 85}, 201101(R) (2012).
\bibitem{SL-1}Y. Shimizu, K. Miyagawa, K. Kanoda, M. Maesato, and G. Saito, Phys. Rev. Lett. {\bf 91}, 107001 (2003).
\bibitem{SL-3}H. Morita, S. Watanabe, and M. Imada, J. Phys. Soc. Jpn. {\bf 71}, 2109 (2002).
\bibitem{SL-2}T. Yoshioka, A. Koga, and N. Kawakami, Phys. Rev. Lett. {\bf 103}, 036401 (2009).
\bibitem{OC-1}I. K$\acute{\rm e}$zsm$\acute{\rm a}$rki {\it et al}., Phys. Rev. B {\bf 74}, 201101(R) (2006).
\bibitem{OC-2}K. Kornelsen, J. E. Eldridge, H. H. Wang, H. A. Charlier, and J. M. Williamst, Solid State Commun. {\bf 81}, 343 (1992).
\bibitem{OC-3}T. Sasaki {\it et al}., Phys. Rev. B {\bf 69}, 064508 (2004).
\bibitem{MTC-exp-V}P. Limelette {\it et al}., Science {\bf 302}, 89 (2003).
\bibitem{MTC-exp-kapp}F. Kagawa, K. Miyagawa, and K. Kanoda, Nature {\bf 436}, 534 (2005).
\bibitem{MI-crit-2}T. Misawa, Y. Yamaji, and M. Imada, J. Phys. Soc. Jpn. {\bf 75}, 083705 (2006).
\bibitem{MI-crit-3}T. Misawa and M. Imada, Phys. Rev. B {\bf 75}, 115121 (2007).
\bibitem{MTC-kapptheo-2}S. Papanikolaou {\it et al}., Phys. Rev. Lett. {\bf 100}, 026408 (2008).
%
\bibitem{MTC-kapptheo-3}M. Sentef, P. Werner, E. Gull, and A. P. Kampf, Phys. Rev. B {\bf 84}, 165133 (2011).
\bibitem{Kubo}R. Kubo, J. Phys. Soc. Jpn. {\bf 12}, 570 (1957).
\bibitem{CDMFT}G. Kotliar, S. Y. Savrasov, G. P$\acute{\rm a}$lsson, and G. Biroli, Phys. Rev. Lett. {\bf 87}, 186401 (2001).
\bibitem{CTQMC}P. Werner, A. Comanac, L. de' Medici, M. Troyer, and A. J. Millis,  Phys. Rev. Lett. {\bf 97}, 076405 (2006).
%
\bibitem{GF-cal-1}T. D. Stanescu and G. Kotliar, Phys. Rev. B {\bf 74}, 125110 (2006).
\bibitem{GF-cal}S. Sakai, Y. Motome, and M. Imada, Phys. Rev. B {\bf 82}, 134505 (2010).
%
\bibitem{GF-cal-2}S. Sakai {\it et al}., Phys. Rev. B {\bf 85}, 035102 (2012).
\bibitem{OC-DMFT}M. J. Rozenberg {\it et al}., Phys. Rev. Lett. {\bf 75}, 105 (1995).
\bibitem{OC-DMFT-2}J. Merino and R. H. McKenzie, Phys. Rev. B {\bf 61}, 7996 (2000).
\bibitem{SC-DMFT}M. Jarrell, Phys. Rev. Lett. {\bf 69}, 168 (1992).
%
\bibitem{SC-DMFT-1}H. Kusunose, J. Phys. Soc. Jpn. {\bf 75}, 054713 (2006).
%
\bibitem{SC-DMFT-2}A. Toschi, A. A. Katanin, and K. Held, Phys. Rev. B {\bf 75}, 045118 (2007).
%
\bibitem{SC-DMFT-3}J. Kune$\check{\rm s}$, Phys. Rev. B {\bf 83}, 085102 (2011).
%
\bibitem{SC-DMFT-4}G. Rohringer, A. Valli, and A. Toschi, Phys. Rev. B {\bf 86}, 125114 (2012).
%
\bibitem{OC-CDMFT-TJ-nov-1}K. Haule and G. Kotliar, Europhys. Lett. {\bf 77}, 27007 (2007).
%
\bibitem{OC-CDMFT-TJ-nov-2}K. Haule and G. Kotliar, Phys. Rev. B {\bf 76}, 104509 (2007).
\bibitem{OC-CDMFT-nov1}S. Chakraborty, D. Galanakis, and P. Phillips, Phys. Rev. B {\bf 78}, 212504 (2008).
%
\bibitem{OC-CDMFT-nov2}N. Lin, E. Gull, and A. J. Millis, Phys. Rev. B {\bf 82}, 045104 (2010).
\bibitem{OC-DCA}N. Lin, E. Gull, and A. J. Millis, Phys. Rev. B {\bf 80}, 161105(R) (2009).
%
\bibitem{OC-DCA-2}N. Lin, E. Gull, and A. J. Millis, Phys. Rev. Lett. {\bf 109}, 106401 (2012).
\bibitem{VC}P. Nozi$\acute{\rm e}$res, {\it Theory of Interacting Fermi Systems} (Addison Wesley, 1964).
\bibitem{MEM}M. Jarrell and J. E. Gubernatis, Phys. Rep. {\bf 269}, 133 (1996).
%
\bibitem{phase}Y. Furukawa: Master's Thesis, Kyoto University, Kyoto (2010).
%
\bibitem{A-1}B. Kyung, Phys. Rev. B {\bf 75}, 033102 (2007).
%
\bibitem{A-2}D. Galanakis, T. D. Stanescu, and P. Phillips, Phys. Rev. B {\bf 79}, 115116 (2009).
%
\bibitem{Ak-1}T. Ohashi, T. Momoi, H. Tsunetsugu, and N. Kawakami, Phys. Rev. Lett. {\bf 100}, 076402 (2008).
\bibitem{JaNa-1}B. Efron and R. J. Tibshirani, {\it An introduction to the bootstrap} (Chapman and Hall/CRC, 1993).


\end{thebibliography}
\end{document}